\documentstyle[aps,prl,epsf]{revtex}
\newcommand{\Sup}{\uparrow}
\newcommand{\Sdn}{\downarrow}
\def\frac#1#2{{#1\over#2}}
\def\dfrac#1#2{{\displaystyle{#1\over#2}}}

\begin{document}
%\voffset=1truecm
\draft
\tighten
\twocolumn[\hsize\textwidth\columnwidth\hsize\csname
@twocolumnfalse\endcsname
\title{Electron Correlations in a Quantum Dot with 
Bychkov-Rashba Coupling}
\author{Tapash Chakraborty$^{\ast\ddag}$ 
and Pekka Pietil\"ainen$^{\ast\dag}$}
\address{$^\ast$Department of Physics and Astronomy,
University of Manitoba, Winnipeg, Canada R3T 2N2}
\address{$^\dag$Departement of Physical Sciences/Theoretical Physics,
P.O. Box 3000, FIN-90014 University of Oulu, Finland} 
\date{\today}
\maketitle
\begin{abstract}
We report on a theoretical approach developed to 
investigate the influence of Bychkov-Rashba 
interaction on a few interacting electrons confined in a
quantum dot. We note that the spin-orbit coupling profoundly
influences the energy spectrum of interacting electrons in 
a quantum dot. Inter-electron interaction causes level
crossings in the ground state and a jump in
magnetization. As the coupling strength is increased, that
jump is shifted to lower magnetic fields. Low-field magnetization 
will therefore provide a direct probe of the spin-orbit 
coupling strength in a quantum dot.
\end{abstract}
\vskip2pc]
\narrowtext   
It has long been recognized that a two-dimensional 
electron gas (2DEG) in narrow-gap semiconductors, 
particularly in InAs-based systems with its high values 
of the $g$-factor, exhibit zero-field splitting
due to the spin-orbit (SO) coupling \cite{alpha}.  
This coupling is also the driving mechanism for making 
futuristic devices based on controlled spin transport, 
such as a spin transistor \cite{dattadas,today}, where the
electron spins would precess (due to the SO coupling)
while being transported through the 2DEG 
channel. Tuning of this precession in the proposed 
spin transistor would provide an additional control 
that is not available in conventional devices, but may
be crucial for the rapidly emerging field of semiconductor
spintronics \cite{today}. Hence the upsurge of interest 
in recent years for a better understanding of the 
SO coupling in nanostructured systems.

The spin-orbit interaction in semiconductor 
heterostructures can be caused by an electric field 
perpendicular to the 2DEG. Riding on an electron, this 
electric field will be {\it felt} as an effective magnetic 
field lying in the plane of the 2DEG, perpendicular to 
the 2D wave vector $k$ of the electron. We consider the
Bychkov-Rashba (BR) Hamiltonian \cite{rashba,note}
$${\mathcal H}_{\rm BR}=
\frac{e \hbar^2}{(2m_0c)^2}\left(
{\vec k}\times{\vec{\bf E}}\right)\cdot{\vec{\bf\sigma}},$$
where $\vec{\bf E}$ is the confining electric field 
at the 2DEG, $\vec{\sigma}=(\sigma_x, \sigma_y, \sigma_z)$
denotes the Pauli spin matrices, and $c$ is the speed of
light. The single-electron Hamiltonian for the 2DEG with 
the electric field normal to the interface,
${\vec{\bf E}}=(0,0,E_z)$, takes the form 
$${\mathcal H}=
-\frac{\hbar^2}{2m^*}\left(\frac{\partial^2}{\partial
x^2}+\frac{\partial^2}{\partial y^2}\right)+ {\rm i}
\alpha\left(\sigma_y\frac{\partial}{\partial x}-\sigma_x
\frac{\partial}{\partial y}\right),$$
where $\alpha$ is the SO coupling parameter
which is sample dependent and is
proportional to the electric field (interfacial and
externally applied). Experimentally observed values of 
$\alpha$ lie in the range of 5 - 45 meV nm \cite{alpha}.
The energy dispersion then consists of two branches
$${\mathcal E}^{\pm}(k)=\frac{\hbar^2}{2m^*}k^2 \pm 
\alpha k$$ with an energy separation $\Delta_{\rm SO}
={\mathcal E}^+ - {\mathcal E}^-=2\alpha k$ for a 
given $k$. The corresponding wave functions are 
\begin{eqnarray*}
\Psi^{\pm}(k_x,k_y)&=& \chi^\pm(k_x,k_y)\,{\rm e}^{
{\rm i}k_xx+{\rm i}k_yy}\\
&=&
\frac1{\sqrt2}\left(\begin{array}{cc}
1 \\ \pm
\dfrac{-{\rm i}k_x+k_y}{k}
\end{array}\right)
{\rm e}^{{\rm i}k_xx+{\rm i}k_yy}
\end{eqnarray*}
The spin parts of the wave functions $\chi^\pm(k_x,k_y)$
are mutually orthogonal and
$\langle\chi^\pm|\sigma_z|\chi^\pm\rangle=0.$
Therefore in the states $\Psi^\pm$ the spins of the 
electrons lie in the $xy$-plane and point in opposite 
directions. In addition, 
$$\langle\chi^\pm|\sigma_x|\chi^\pm\rangle=\frac{2k_y}k, 
\qquad
\langle\chi^\pm|\sigma_y|\chi^\pm\rangle=-\frac{2k_x}k,$$
i.e., the spins are {\it perpendicular} to the momentum 
$(k_x,k_y)$. Spatial alignment of spins therefore depends 
on the wave vector \cite{alpha}.

Spin-orbit interaction and electron-electron interactions are  
responsible for a variety of interesting effects in 
quantum dots \cite{bert}. In this paper, we present a numerically
exact treatment of the BR Hamiltonian in a system of interacting 
electrons confined in a parabolic quantum dot 
(QD) \cite{makchak,qdbook} under the influence of an external 
magnetic field. More specifically, we explore the energy spectra 
and magnetization \cite{singtrip} of a few interacting electrons 
in a quantum dot in the presence of SO coupling. It should 
be pointed out that while a large number of theoretical work 
has been reported as yet in the literature for a 2DEG 
\cite{single} and a QD \cite{qdot_kuan,qdot} with spin-orbit 
coupling, in most cases, the electron-electron interaction has been 
ignored (or treated within an approximation \cite{coulomb}) due to 
its inherent complexity. 

Let us begin with the single-electron states. Unlike the case of 
a circular quantum dot with hard walls where exact analytical results 
for the single-electron energy spectrum (in the presence of the 
SO interaction) are available, for the more realistic
case of a parabolic QD, the energy spectrum can only be obtained 
numerically. In the presence of the BR interaction, the Schr\"odinger 
equation consiste of two parts
\begin{eqnarray}
\nonumber
-\frac{\hbar^2}{2m^*}\nabla^2\psi^\Sup+\alpha\nabla^-\psi^\Sdn+
v_c({\bf r})\psi^\Sup&=&\varepsilon\psi^\Sup\\
-\frac{\hbar^2}{2m^*}\nabla^2\psi^\Sdn-\alpha\nabla^+\psi^\Sup+
v_c({\bf r})\psi^\Sdn&=&\varepsilon\psi^\Sdn\\
\nonumber
\end{eqnarray}
where $\psi$ is a two-component spinor
\begin{equation}
\psi=\left(\begin{array}{c}
\psi^\Sup\\
\psi^\Sdn\\
\end{array}\right),
\end{equation}
$\nabla^\pm=\frac{\partial}{\partial x}\pm 
i\frac{\partial}{\partial y}$,
and $v_c({\bf r})$ is the confinement potential. We seek a solution
of the form
\begin{eqnarray*}
\psi^\Sup &=& f_\Sup(r)\,e^{im_\Sup\theta}\\
\psi^\Sdn &=& f_\Sdn(r)\,e^{im_\Sdn\theta}
\end{eqnarray*}
which with Eq.~(1) yields, $m_\Sup=m_\Sdn-1 (=m).$
In the case of a parabolic confinement potential
$v_c=\frac12m^*\omega_0^2r^2$, the radial equations are
\begin{eqnarray}
\nonumber
xf_\Sup^{''}+f'_\Sup&+&\left(\nu-\frac{m^2}{4x}-\frac x4\right)f_\Sup\\
&-&\beta x^\frac12\left(f'_\Sdn+\frac{m+1}{2x}f_\Sdn\right)=0\\
\nonumber
xf_\Sdn^{''}+f'_\Sdn&+&\left(\nu-\frac{(m+1)^2}{4x}-\frac x4\right)f_\Sdn\\
\nonumber
&+&\beta x^\frac12\left(f'_\Sdn-\frac m{2x}f_\Sup\right)=0\\
\nonumber
\end{eqnarray}
where $x=r^2/a^2$, $a^2=\hbar/(m^\ast\omega_0)$,
$\nu=\varepsilon/(2\hbar\omega_0)$ and $\beta=m^\ast a\alpha/\hbar^2$.
When $\beta=0$ (i.e., $\alpha=0$), Eq.~(3) reduces to two uncoupled 
Laguerre equations with solutions
\begin{eqnarray*}
f_\Sup &=& f_{nm}=e^{-x/2}x^{\vert m\vert/2}L_n^{\vert m\vert}\\
f_\Sdn &=& f_{n,m+1}
\end{eqnarray*}
with the energies,
\begin{equation}
\nu_{nm}=n+\frac{|m|+1}2.
\end{equation}
In the presence of an external magnetic field $B$ the term
$$
{\cal H}_B=
\frac{e^2B^2r^2}{8m^\ast c^2}
-\frac{ieB\hbar}{2m^\ast c}\frac{\partial}{\partial\theta}
+\frac{e\alpha B}{2\hbar c}r
\left(
\begin{array}{cc}0&e^{-i\theta} \\ e^{i\theta}&0\end{array}
\right)
+\frac12 g\mu_B B\sigma_z
$$
has to be added to the spinor Hamiltonian $\cal H$. Here
the first two terms (diagonal)
are due to the interaction of the orbital motion
and the magnetic field. The third (non-diagonal) term originates
from the vector potential part $\vec A=\frac B2\left(-y,x,0\right)$ 
in minimal coupling scheme
$\alpha/\hbar\,[\vec\sigma\times(\vec p-e/c\vec A)]_z$
of the SO interaction. The last term gives the Zeeman energies
of the components of the spinors. When $\beta = 0$
the functions $f_{nm}$ will still
be eigenstates of the Hamiltonian provided
that we replace the single particle energies $\nu_{nm}$
with the expressions
$$
\nu_{nm}^\sigma=n+\frac{|m|+1}2-\frac{\omega_c}{4\Omega}m
+\sigma\frac{g\mu_B B}{4\hbar\Omega},
$$
where signs $\sigma=\pm1$ correspond to the upper and lower components of the
spinor. Furthermore, for the angular velocity $\omega_0$ related to
the harmonic confinement
potential we have to substitute the effective angular velocity
$\Omega=\omega_0\left(1+\omega_c^2/(4\omega_0^2)\right)^{1/2}$ where
$\omega_c=eB/(m^\ast c)$ is the cyclotron frequency.

\begin{figure}
\begin{center}
\begin{picture}(120,130)
\put(0,0){\includegraphics{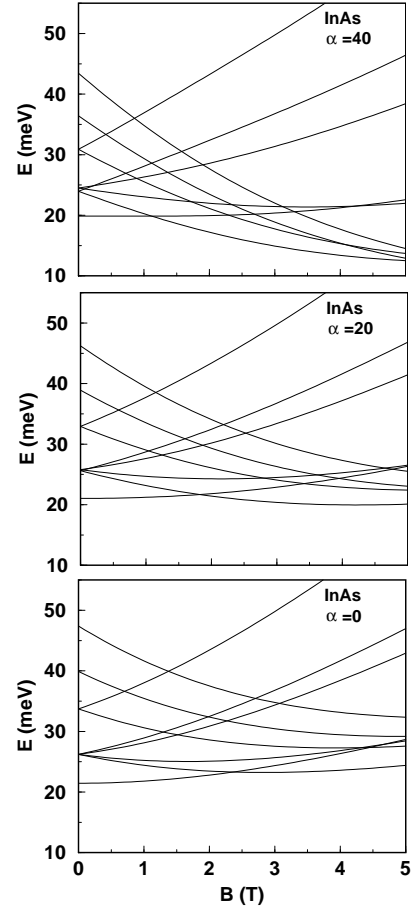}}
\end{picture}
\vspace*{9.0cm}
\caption{Energy spectrum of a two-electron InAs quantum dot
versus the applied magnetic field for different values of
the Bychkov-Rashba interaction parameter $\alpha$ (meV nm). For
clarity, at each value of the magnetic field and for a given
total angular momentum, only the lowest 
energy is plotted. The Zeeman energy is also included.
}
\end{center}
\end{figure}

For the $\beta\not=0$ case which is our main concern here,
we use the following expansion
$$
f_\uparrow=\sum_{n=0}^\infty c_n^\uparrow f_{nm};
\quad
f_\downarrow=\sum_{n=0}^\infty c_n^\downarrow f_{n,m+1}.
$$
For the angular momenta $m\geq0$ we find that
the expansion coefficients satisfy the equations
\cite{qdot_kuan}
\begin{eqnarray*}
(\nu -\nu^\uparrow_{nm})c_n^\uparrow
&=& \frac{\beta}2\left[
\eta^+(n+m+1)c_n^\downarrow + \eta^-nc_{n-1}^\downarrow
\right] \\
(\nu -\nu^\downarrow_{n,m+1})c_n^\downarrow
&=& \frac{\beta}2\left[
\eta^+c_n^\uparrow + \eta^-c_{n+1}^\uparrow
\right]
\end{eqnarray*}
and the equations
\begin{eqnarray*}
(\nu -\nu^\uparrow_{nm})c_n^\uparrow
&=& -\frac{\beta}2\left[
\eta^-c_n^\downarrow + \eta^+c_{n+1}^\downarrow
\right] \\
(\nu -\nu^\downarrow_{n,m+1})c_n^\downarrow
&=& -\frac{\beta}2\left[
\eta^-(n-m)c_n^\uparrow + \eta^+nc_{n-1}^\uparrow
\right]
\end{eqnarray*}
for states with $m<0$, and $\eta^\pm=1\pm ea^2B/(\hbar c)$.
Solutions of these eigensystems provide the single-electron 
energies $\varepsilon$ and the spinor wavefunctions $\psi$ 
[Eq.~(2)], which have been investigated
earlier by several authors in a variety of ways \cite{qdot}.

\begin{figure}
\begin{center}
\begin{picture}(120,130)
\put(0,0){\includegraphics{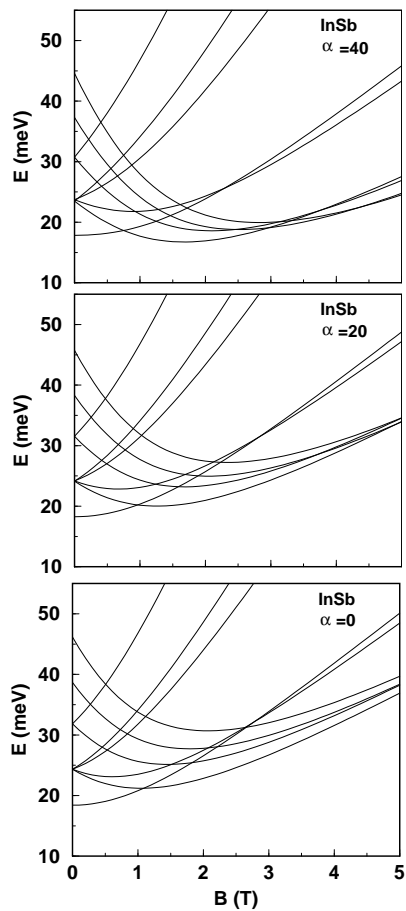}}
\end{picture}
\vspace*{8.0cm}
\caption{Same as in Fig.~1, but for the InSb quantum dot
system.
}
\end{center}
\end{figure}

For a system of interacting electrons we diagonalize
the many-body Hamiltonian in a basis consisting of
non-interacting many-body states, which in turn are constructed
as antisymmetrized direct products of the two-component spinors $\psi$.
Since the Coulomb force is independent of the spin orienation we  
evaluate the sum
$$
\langle\psi_{\lambda_1}\psi_{\lambda_2}|V_{\mbox{\scriptsize Coul}}|
\psi_{\lambda_3}\psi_{\lambda_4}\rangle
=\sum_{\sigma_1,\sigma_2}
\langle\psi_{\lambda_1}^{\sigma_1}\psi_{\lambda_2}^{\sigma_2}
|V_{\mbox{\scriptsize Coul}}|
\psi_{\lambda_3}^{\sigma_2}\psi_{\lambda_4}^{\sigma_1}\rangle,
$$
$\lambda=(n,m)$, of four terms. Explicit expression for these terms for a 
parabolic QD can be found in Ref.~\cite{qdbook}.

In our numerical investigations, we choose InAs and InSb quantum
dots with parameters, $m^*/m_0=0.042, \epsilon=14.6, g=-14$
and $m^*/m_0=0.014, \epsilon=17.88, g=-40$ respectively. InSb quantum dots
are also considered here because of its high $g$ values. In both systems,
we choose $\hbar\omega_0=7.5$ meV. The energy spectrum of the two-electron
state in InAs QD is shown in Fig.~1 for various values
of the BR coupling parameter $\alpha$. Similar results for InSb QD 
are presented in Fig.~2.

The essential feature of the energy spectra at $\alpha=0$ is
that with the increase in magnetic field, the ground state moves 
from $J=0$ to $J=2$ $(J=m+s_z$ is the total angular momentum). This 
is already well established in the literature \cite{singtrip}. This 
level crossing persists for nonzero value of $\alpha$,
but the crossing point shifts to lower magnetic fields.
This shift of the crossing point can 
perhaps be observed experimentally by a variety 
of ways, such as capacitance spectroscopy, or by transport 
spectroscopy \cite{qdbook}.

The results for magnetization at the ground state, defined 
as $M=-\partial E/\partial B$, where $E$ is the total energy
of the system, of quantum dots with or
without the BR interaction is presented in Fig.~3.
Magnetization is a fundamental thermodynamic quantity that
reflects the change of the ground state electron energy in a
magnetic field \cite{magnetize_expt}, thereby providing 
valuable information about many-electron dynamics of the QD 
in a magnetic field. We have established earlier that 
oscillations in magnetization in a few electron-quantum dot 
are a direct consequence of the effects related to the 
electron-electron interaction between the two-dimensional 
electrons confined in the dot \cite{singtrip}. A jump in $M$
occurs at a magnetic field where the ground state changes
from one angular momentum 
to another (Figs.~1, 2). Similar behavior is also expected in 
a nanoscopic quantum ring \cite{nanoring}. With increasing 
strength of $\alpha$, this jump in magnetization at the 
energy-level crossing is pushed to lower magnetic fields. For
the InAs QD this shift can be as large as $\sim1.5$ Tesla when
$\alpha$ is increased from zero to 40 meV nm. Therefore,
low-field magnetization measurements of quantum dots could be
a direct probe of the SO coupling strength.

In closing, we have developed a theoretical approach where
the SO interaction is treated via exact diagonalization of the
Hamiltonian for interacting electrons confined in a parabolic
QD. Coulomb interaction causes energy levels to cross and
at the crossing point magnetization shows a jump. In a magnetic 
field the strength of the SO coupling is proportional
to the field (in addition to the coupling parameter and the
angular momentum). Hence, the effect of the coupling is more 
prominent for slopes of the higher angular momenta energy curves.
As a consequence, an increase in the SO coupling strength 
causes the energy level crossings to move to weaker fields and 
the jump in magnetization shows a large shift to weaker 
magnetic fields. This result can be exploited to tune the SO 
coupling strength that might be useful for spin transport. Our 
theoretical approach can be extended to include larger number 
of electrons in the dot. Details will be published elsewhere.

\begin{figure}
\begin{center}
\begin{picture}(120,130)
\put(0,0){\includegraphics{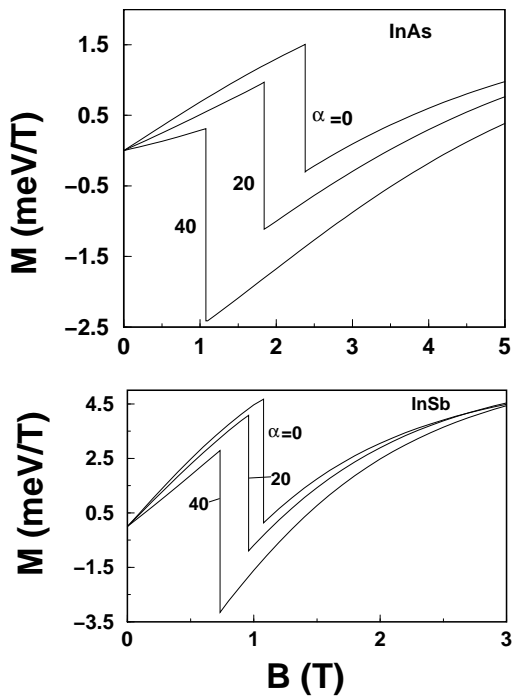}}
\end{picture}
\vspace*{6.0cm}
\caption{Magnetization in the ground state for various values of 
the SO coupling strength and for the two electron QD systems.
}
\end{center}
\end{figure}

We would like to thank Marco Califano for helpful
discussions. The work of T.C. has been supported by the Canada 
Research Chair Program and the Canadian Foundation for Innovation 
Grant.

\end{document}